\documentclass[
amsmath,amssymb,
aps,
prl,
twocolumn,
superscriptaddress,10pt
]{revtex4-2}

\usepackage{graphicx}% Include figure files
\usepackage{dcolumn}% Align table columns on decimal point
\usepackage{bm}% bold math
\usepackage{hyperref}% add hypertext capabilities
\graphicspath{{figures/}}
\usepackage{xcolor}
\usepackage{braket}   
\usepackage{physics}  % physics 宏包也提供 
\usepackage{caption} 
\usepackage{array}
\usepackage{booktabs}  
\usepackage{ragged2e}   
\usepackage[labelsep=period]{caption}  % 放在导言区（导入一次即可）
\usepackage{times} 

\begin{document}
	
	\preprint{APS/123-QED}
	
	\title{\textbf{%Equivalence Classification of MUBs Based on Finite Groups
    %Efficient Classification of Inequivalent MUB Subsets via Finite Operations
    Efficient Identification the Inequivalence of Mutually Unbiased Bases via Finite Operators
    } 
	}% 
\author{Jianxin Song}
\affiliation{Key Laboratory of Low-Dimension Quantum Structures and Quantum Control of Ministry of Education, Synergetic Innovation Center for Quantum Effects and Applications, Xiangjiang-Laboratory and Department of Physics, Hunan Normal University, Changsha 410081, China}
    \affiliation{Hunan Research Center of the Basic Discipline for Quantum Effects and Quantum Technologies, Hunan Normal University, Changsha 410081, China}
	\affiliation{Institute of Interdisciplinary Studies, Hunan Normal University, Changsha 410081, China}	
    
\author{Zhen-Peng Xu}\thanks{Corresponding author: zhen-peng.xu@ahu.edu.cn}
\affiliation{School of Physics, Anhui University, 230601 Hefei, China }

\author{Changliang Ren}\thanks{Corresponding author: renchangliang@hunnu.edu.cn}
\affiliation{Key Laboratory of Low-Dimension Quantum Structures and Quantum Control of Ministry of Education, Synergetic Innovation Center for Quantum Effects and Applications, Xiangjiang-Laboratory and Department of Physics, Hunan Normal University, Changsha 410081, China}
    \affiliation{Hunan Research Center of the Basic Discipline for Quantum Effects and Quantum Technologies, Hunan Normal University, Changsha 410081, China}
	\affiliation{Institute of Interdisciplinary Studies, Hunan Normal University, Changsha 410081, China}

%	\date{\today}

	\begin{abstract}
The structural characterization of high-dimensional mutually unbiased bases (MUBs) by classifying MUBs subsets remains a major open problem. The existing methods not only fail to conclude on the exact classification, but also are severely limited by computational resources and suffer from the numerical precision problem. Here we introduce an operational approach to identify the inequivalence of MUBs subsets, which has less time complexity and entirely avoids the computational precision issues. For arbitrary MUBs subsets of $k$ elements in any prime dimension, this method yields a universal analytical upper bound for the amount of MUBs equivalence classes. By applying this method through simple iterations, we further obtain tighter classification upper bounds for any prime dimension $d\leq 37$. Crucially,  the comparison of these upper bounds with existing lower bounds successfully determines the exact classification for all MUBs subsets in any dimension $d \leq 17$. We further extend this method to the case that the dimension is a power of prime number. This general and scalable framework for the classification of MUBs subsets sheds new light on related applications.
	\end{abstract}
	
	\maketitle
%	\tableofcontents
		
	\textit{Introduction.---} 
    Incompatible measurements are the cornerstones of quantum information theory and techniques. As the “maximally incompatible" measurements\cite{Schwinger.pnas.46.4.570,Ivonovic.J.Phys.A.14.3241,Kraus.PhysRevD.35.3070,William.Ann.Phys.191.363}, 
    mutually unbiased bases ($\text{MUBs}$) plays a crucial role in quantum tasks such as quantum state tomography  \cite{NicolasJ.PhysRevLett.88.127902,Sosa.PhysRevLett.119.150401,Bent.PhysRevX.5.041006,Adamson.PhysRevLett.105.030406,Jaroslav.PhysRevA.92.052303,Fern.PhysRevA.83.052332,Parthasarathy.QPRT.7,Bendersky.PhysRevLett.100.190403,Fano.Rev.Mod.Phys.29.74}, quantum key distribution \cite{Armin.Sci.Adv.7.eabc3847,Cerf.PhysRevLett.88.127902,Ikuta.PhysRevResearch.4.L042007,Mafu.PhysRevA.88.032305,Sheridan.PhysRevA.82.030301,Bartkiewicz.PhysRevA.93.062345}, quantum random number generation \cite{Bechmann.PhysRevLett.85.3313,Cao.PhysRevX.6.011020,Ma.PhysRevA.99.022328} and detection of quantum nonlocality \cite{Siudzińska.Sci.Rep.11.22988,Liu.Sci.Rep.5.13138,Giovannini.PhysRevLett.110.143601,Spengler.PhysRevA.86.022311,Aharonov.PhysRevLett.47.1029,Czartowski.PhysRevLett.124.090503,Tavakoli.RevModPhys.96.045006}. 
    
    In a $d$-dimensional Hilbert space, two orthonormal bases $\Set{\Ket{e_i}}$ and $\Set{\Ket{f_j}}$ are called MUBs if $|\langle e_i | f_j \rangle|^2 = 1/d$ for all $i, j$. 
    One main research line on this topic is to understand the structure of MUBs, such as determining the maximum amount of MUBs in a given dimension. Such a question has been solved constructively for the prime-power dimension $d$, where the the maximum amount of MUBs is $d+1$. However, even when $d$ is the simple composite number $6$, it is still an open question after decades of research~\cite{Brierley.PhysRevA.79.052316,Durt.J.Phys.A.38.5267,Philippe.Crypt.Comm.2.211,Ambrosio.Sci.Rep.3.2726,Lawrence.PhysRevA.65.032320}. In Parallel, a new perspective to explore the MUB structures is from determining the inequivalence classes of subsets of $k$ MUBs.
    %The entirety of these bases constitutes a maximum complete MUB set.
Two subsets of $k$  $\text{MUBs}$ are equivalent if and only if one subset can be transformed into another one by operations including unitary rotations, complex conjugation, global phases, column permutations, and reordering of bases \cite{Brierley.QuantumInf.Comput.10.0803,Daniel.arxiv.2410.23997,Kraus.PhysRevLett.104.020504,Kraus.PhysRevA.82.032121}. In high dimensions ($d\ge 5$), MUBs subsets are known to exhibit multiple inequivalence classes \cite{Yan.PhysRevLett.132.080202,Laura.Phys.Rev.Res.7.033152,Designolle.PhysRevLett.122.050402,Designolle.PhysRevA.105.032430,Aguilar.PhysRevLett.121.050501,Hiesmayr.New.J.Phys.23.093018,Tendick.PhysRevLett.131.120202}. While, direct classification by definition is impractical, as it requires exploring an uncountable set of operations. To address this, the operational criterion based on Shannon entropy \cite{Deutsch.PhysRevLett.50.631,Maassen.PhysRevLett.60.1103,Riccardi.PhysRevA.95.032109,Trifonov.J.Phys.A.31.8041,Laura.Phys.Rev.Res.7.033152,Coles.RevModPhys.89.015002,Wehner.New.J.Phys.12.025009,Wu.PhysRevA.79.022104,Ballester.PhysRevA.75.022319},  robustness of incompatible measurements \cite{Designolle.PhysRevLett.122.050402,Uola.RevModPhys.95.011003,Designolle.PhysRevA.105.032430}, accuracy of quantum random access code \cite{Aguilar.PhysRevLett.121.050501}, entanglement detection efficiency  \cite{Hiesmayr.New.J.Phys.23.093018}, and diamond distance  \cite{Tendick.PhysRevLett.131.120202,Thomas.arxiv.2207.05722,Tendick.Quantum.7.1003} have been proposed.
Since those criteria are only sufficient conditions for the inequivalent classes, 
they generally provide only the lower bounds of the amount of classes in the classification. 
%Especially, different criteria have different detection power, e.g., . 
It is not known whether any criterion exactly characterize the classification.
More importantly, as the dimension of the system or the size of the subsets increases, those numerical methods encounter the following obstacles: (i)  optimization procedures easily get trapped in local optima instead of the global one, e.g., the criterion based on Shannon entropy is a non-convex programming; (ii) the distinction between the values of different classes are so subtle that the resulting classification might be spurious; (iii) the requirement of computational resources goes beyond the available capacity quickly as the dimension of the system and the size of subsets increases.

In this work, we proposes an analytical method to classify MUBs subsets based on finite transformation group. By using a subset of those operations, we first derive a universal upper bound for the amount of inequivalent classes of MUBs subsets in any prime dimension $d$, which yields the exact classification for $d=5$. A further systematical application of this method leads to a more tight upper bound on the amount of inequivalent classes for any other prime dimension $d \le 37$. Critically, such an upper bound coincides with the lower bound derived from the Shannon entropy criterion, and thus successfully determines the exact classification for all MUBs subsets of any size $k$ in prime dimension $d \leq 17$, and cases of small $k$ for higher dimension $d$. The method is also generalized for prime-power dimensions by incorporating more transformations, which results in the exact classification for dimension $d=9$. Finally, complexity analysis and practical performance reveal that this approach is computationally lighter than the existing methods.

   \textit{ \text{MUBs classification using finite unitary matrices.---} }In a Hilbert space of prime dimension $d$, there exists a complete set of $d+1$ $\text{MUBs}$ \cite{Schwinger.pnas.46.4.570,Ivonovic.J.Phys.A.14.3241},
 %  which can be defined by 
   denoted as $\mathcal{M} = \{M_0, \dots, M_d\}$. Any  element $M_x$ in the set of $\text{MUBs}$  is composed of $d$ complete orthonormal basis vectors $\{\ket{\psi_{a,x}}\}$,
 %Each $M_x$ can be specified by a complete set of $d$ orthonormal basis vectors $\{\ket{\psi_{a,x}}\}$, \textcolor{red}{Note:Is it a bit repetitive?} 
 which can be defined as 
    \begin{align}
        \ket{\psi_{a,x}} &= 
            \frac{1}{\sqrt{d}}
            \sum_{j=0}^{d-1} 
            \omega^{x j^{2} - a j}\ket{j}, 
            && x \in \{0,1,\dots,d-1\},\notag \\
        \ket{\psi_{a,d}} &= \sum_{j=0}^{d-1}\delta_{j,a}\ket{j}, 
            && x = d,
    \end{align}
    where $\ket{j}$ is computational basis, $\omega = e^{2\pi i/d}$, and $a \in \{0,1,\dots,d-1\}$. The main task is to determine the exact amount of inequivalence classes formed by arbitrarily selected subsets of $k$ $\text{MUBs}$ from this collection for a give positive integer $k \le d+1$.
   
    Since unitary transformations are arbitrary, it is generally infeasible to distinguish inequivalent MUBs subsets directly from their definition. We define a finite set of unitary transformation matrices containing only $d+1$ elements based on the MUBs set $\mathcal{M}$, denoted as $\mathcal{U}_{\mathrm{finite}} = \{M_0, \dots, M_d\}$. 
   We rigorously prove that any operation $M_x$ within this finite unitary set $\mathcal{U}_{\mathrm{finite}}$ possesses a crucial closure property: it transforms the orthonormal basis vectors $\ket{\psi_{b,y}}$ of any $y$-th MUB into the basis vectors $\ket{\psi_{c,z}}$ of the $z$-th MUB within the complete set $\mathcal{M}$. This establishes a one-to-one correspondence ($y \to z$) among the MUB indices for a fixed $M_x$.  Specifically, when $x = d$, $M_d$ is the identity matrix, and the basis vectors $\{\ket{\psi_{b,y}}\}$ remain unchanged. For $x \in \{0,1,\dots,d-1\}$ and $y \in \{1,\dots,d-1\}$, the new basis vector indices satisfy the exact analytic relations $\{z = x - \frac{1}{4y} \pmod{d}, \quad c = \frac{b}{2y} \pmod{d}\}$. Specifically, when $y = 0$, any $M_x$ transforms the basis vectors $\{\ket{\psi_{b,0}}\}$ into the computational basis vectors. When $y = d$, the unitary operator $M_x$ transforms $\{\ket{\psi_{b,d}}\}$ into the orthonormal basis vectors corresponding to $M_x$ itself. This explicit, finite, and closed mapping is central to our MUBs classification framework (detailed proof in Sec.II of Supplemental Material(SM)).

   Based on this finite set of unitary operations and their tractable transformation, we derive a  universal analytical upper bound for the amount of MUBs inequivalence classes. \\
   % \textit{\textbf{Theorem 1.}} \textit{In a prime-dimensional Hilbert space, if $k=1$ or $k=d$, there exists only one equivalence class of MUBs subsets; for $1 < k < d$, the amount of inequivalent classes of any MUBs subset formed by $k$ sets is at most $\binom{d+1}{k} / {2d}$.}
 \textit{\textbf{Theorem 1.}}\textit{ In a prime-dimensional Hilbert space of dimension $d$, the classification of mutually unbiased bases (MUBs) subsets satisfies the following properties:
\begin{align}
N_{k} \leq 
\begin{cases}
\quad 1, & \text{if }\quad k = 1 \text{ or }  d,\\[6pt]
\dfrac{\binom{d+1}{k}}{2d}, & \text{if }\quad 1 < k < d.
\end{cases}
\end{align}
}

    Since there is at least one class, the derived upper bound is saturated in the case that $k = 1$ or $k = d$.  For $1 < k < d$, the upper bound is derived by proving that the finite unitary group $\mathcal{U}_{\mathrm{finite}}$, acting on the complete MUBs sets $\mathcal{M}$, generates a minimum of $2d$ inequivalent subsets (orbits). Details of the proof are provided in Sec.III of SM. 
    Crucially, for the first known case of existing multiple inequivalent classes ($d=5$, $k=3$) \cite{Designolle.PhysRevLett.122.050402},  this upper bound exactly yields $N_{\mathrm{UB}} = \binom{6}{3}/{10} = 2$, perfectly matching the known amount of  inequivalent classes \cite{Yan.PhysRevLett.132.080202,Laura.Phys.Rev.Res.7.033152,Designolle.PhysRevLett.122.050402,Designolle.PhysRevA.105.032430,Aguilar.PhysRevLett.121.050501,Hiesmayr.New.J.Phys.23.093018,Tendick.PhysRevLett.131.120202}. However, the tightness of this bound deteriorates rapidly in higher dimensions due to the further overlap of distinct orbits. To achieve a tighter upper bound, we refine our method by incorporating complex conjugation operations into the $\mathcal{U}_{\mathrm{finite}}$ group. Most notably, combining this tighter upper bound with existing lower bounds~\cite{Designolle.PhysRevLett.122.050402} allows us to achieve, for the first time, the exact classification of MUBs inequivalence classes for dimensions  $d \leq 17$.

\begin{table}[htbp]
        \centering
        \captionsetup{
            justification=justified,
            singlelinecheck=false
        }
        \caption{\label{tab:mub_d5_transform}%
        \justifying
     Transformation table of the finite operations in $d=5$  }
        \begin{ruledtabular}
        \begin{tabular}{c|cccccc}
                $U \backslash M_y$ & $M_0$ & $M_1$ & $M_2$ & $M_3$ & $M_4$ & $M_5$ \\
                \hline
                $M_0$ & 5 & 1 & 3 & 2 & 4 & 0 \\
                $M_1$ & 5 & 2 & 4 & 3 & 0 & 1 \\
                $M_2$ & 5 & 3 & 0 & 4 & 1 & 2 \\
                $M_3$ & 5 & 4 & 1 & 0 & 2 & 3 \\
                $M_4$ & 5 & 0 & 2 & 1 & 3 & 4 \\
                $M_5$ & 0 & 1 & 2 & 3 & 4 & 5 \\
                $conj$ & 0 & 4 & 3 & 2 & 1 & 5 \\
            \end{tabular}
        \end{ruledtabular}
    \end{table}
    
    \begin{figure*}[htbp]
        \centering
         \includegraphics[width=0.8\textwidth]{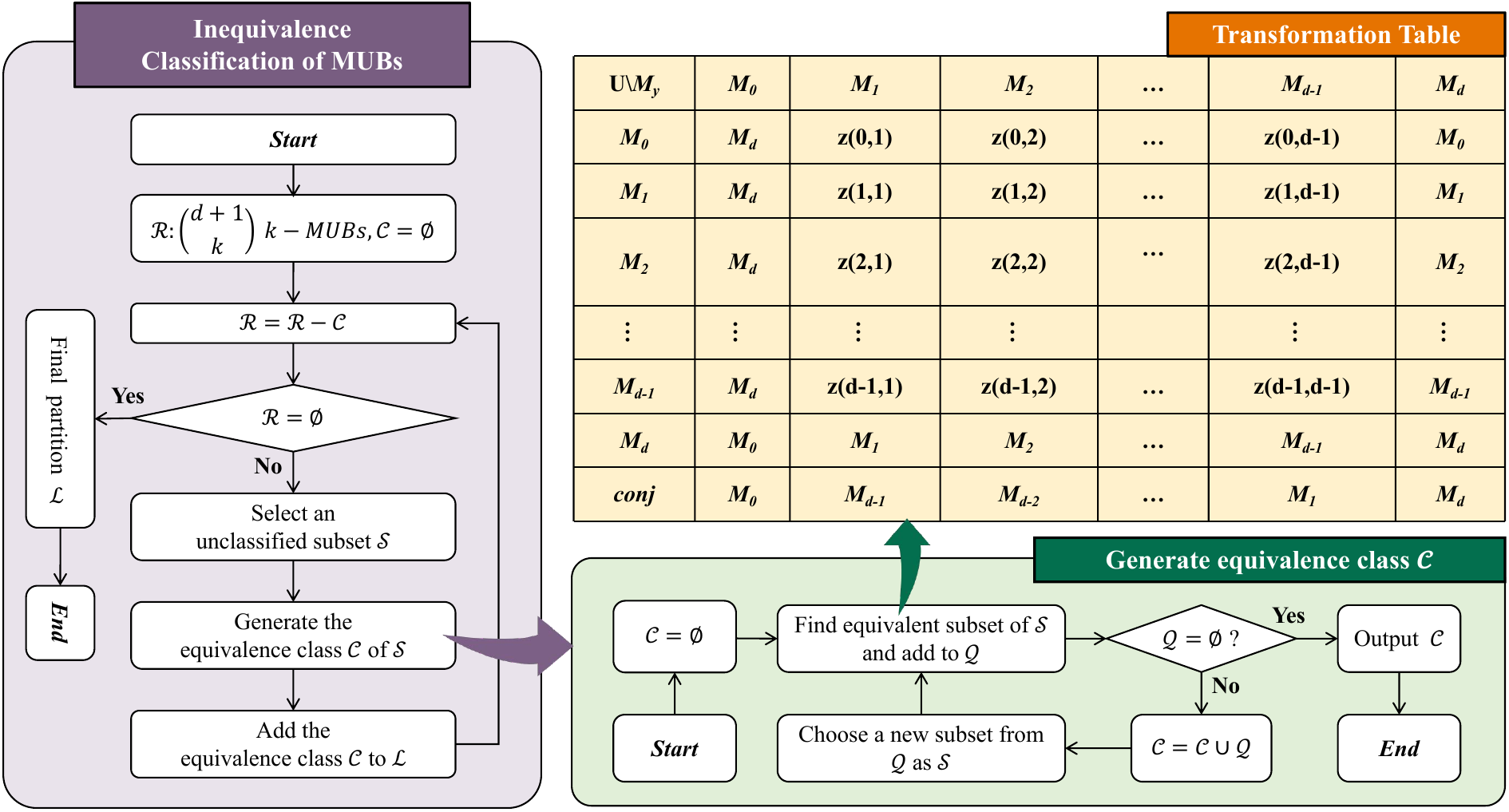}
        \caption{\justifying Schematic of the finite-group classification method. $\mathcal{R}$, $\mathcal{Q}$ and $\mathcal{L}$ represent the unclassified collection, the processing queue  and the  classified collection, respectively.}
        \label{Classification_schematic}
    \end{figure*}

  \textit{ Determining the Exact Classification for Prime Dimensions based on Upper and Lower Bounds.---} Our core methodology determine MUBs inequivalence classes by analyzing transformations under the finite unitary group $\mathcal{U}_{\mathrm{finite}}$ and complex conjugation. Complex conjugation maintains operational equivalence, as any basis vector $\ket{\psi_{b,y}}$ transforms analytically to a vector within the complete set $\mathcal{M}$ via the relation $z = -y \pmod{d}, c = -b \pmod{d}$. This approach is significant because it not only provides bounds on the amount of inequivalent classes but, more importantly, enables the precise identification of their sets.
  
  %The core method adopted aims to determine equivalent MUBs sets and provide an upper bound on the number of inequivalent MUBs sets by analyzing the equivalence of MUBs collections under the action of finite unitary operations ($\mathcal{U}_{\mathrm{finite}}$) and complex conjugation transformations. This approach not only analyzes the number of inequivalent classes but, more importantly, can precisely identify and determine the specific sets of these equivalence classes. Here, the complex conjugation operation corresponds to MUBs equivalence: any basis vector $\ket{\psi_{b,y}}$ remains within the full set $\mathcal{M}$ after complex conjugation, with the transformation relation $z = -y \pmod{d}, \quad c = -b \pmod{d}$. If two MUBs subsets can be completely transformed through the action of $\mathcal{U}_{\mathrm{finite}}$ and complex conjugation, they are determined to be equivalent.

    \textit{MUBs Inequivalence Classes Classification Method.---} As shown in Fig.~\ref{Classification_schematic}, this method includes the following three steps:

 (1). By traversing all unitary transformations in $\mathcal{U}_{\mathrm{finite}}$ and their complex conjugates, we construct a $(d+2)\times(d+1)$ ``Transformation Table'', which records the action on each $M_y \in \mathcal{M}$ and enables efficient indexing for equivalence analysis.
 
 %Traverse all unitary transformations in $\mathcal{U}_{\mathrm{finite}}$ and their complex conjugates, and for each $M_x \in \mathcal{M}$ record the resulting transformations to obtain a $(d+2) \times (d+1)$ “Transformation Table,” which enables rapid indexing for subsequent equivalence analysis.

 (2). Starting with an initial subset $\mathcal{S} = \{M_1, \dots, M_k\}$, we determine its complete equivalence class $\mathcal{C}$ via an iterative search. The $d+2$ equivalent subsets are rapidly identified using the transformation table. Each newly generated subset is subjected to standardized ordering (ascending by MUB index) to ensure unique representation, stored in the set $\mathcal{C}$ (removing duplicates), and added to the processing queue $\mathcal{Q}$. This procedure is repeated iteratively by selecting subsets from $\mathcal{Q}$ until the queue is exhausted ($\mathcal{Q} = \varnothing$). Then $\mathcal{C}$ fully contains all subsets equivalent to the initial set $\mathcal{S}$.

 %Generation of Equivalence Classes — Select any $k$ bases to form an initial subset $\mathcal{S} = \{M_1, \dots, M_k\}$, and construct the equivalence class $\mathcal{C}$ containing all its equivalent subsets. From the constructed transformation table, extract the $d+2$ subsets that can be completely transformed under the equivalence operations, and perform a standardized ordering (ascending by MUBs index) for each newly generated subset to ensure uniqueness of representation. Store all standardized subsets in the set $\mathcal{C}$ (removing duplicates) and add any newly identified subsets to a processing queue $\mathcal{Q}$. This procedure is iteratively continued until $\mathcal{Q}$ is empty. At this point, $\mathcal{C}$ contains all subsets obtainable from $\mathcal{S}$ via the equivalence operations, i.e., $\mathcal{C}$ constitutes the complete equivalence class of $\mathcal{S}$.

(3). Defining the initial set $\mathcal{R}$ as the complete collection of $\binom{d+1}{k}$ $k$-MUBs subsets, the classification proceeds by iteratively selecting an unclassified subset $\mathcal{S}$ from $\mathcal{R}$, generating its complete equivalence class $\mathcal{C}$, and adding $\mathcal{C}$ to the final partition $\mathcal{L}$. All elements of $\mathcal{C}$ are then removed from $\mathcal{R}$. This procedure is repeated until $\mathcal{R} = \varnothing$, yielding $\mathcal{L}$ as the complete classification of all inequivalent $k$-MUBs subsets.

%Partitioning the Full Set — Define the initial set $\mathcal{R}$ to be classified as the collection of all $\binom{d+1}{k}$ $k$ bases subsets. Select any unclassified subset $\mathcal{S}$ from $\mathcal{R}$, determine its complete equivalence class $\mathcal{C}$ using the aforementioned equivalence class generation mechanism, and remove all elements of $\mathcal{C}$ from $\mathcal{R}$. Repeat this procedure until the set $\mathcal{R}$ to be classified is empty ($\varnothing$). The final output set $\mathcal{L}$ then constitutes the partition of inequivalent $k$ MUBs subsets.

As an illustration, consider $d=5$, where $\mathcal{U}_{\text{finite}}=\{M_0,\dots,M_5\}$.  
Applying each $U\in\mathcal{U}_{\mathrm{finite}}$ and its conjugate to the full set $\mathcal{M}$ yields the complete ``Transformation Table'' (Table~\ref{tab:mub_d5_transform}). This enables efficient classification: for $k=2$, all $\binom{6}{2}=15$ subsets belong to a single equivalence class; for $k=3$, two inequivalent classes emerge (e.g., $\{M_0,M_1,M_2\}$ and $\{M_0,M_1,M_3\}$). These results fully agree with previous findings~\cite{Laura.Phys.Rev.Res.7.033152,Designolle.PhysRevLett.122.050402,Designolle.PhysRevA.105.032430,Aguilar.PhysRevLett.121.050501,Hiesmayr.New.J.Phys.23.093018}.  Extending this method, we determine both upper bounds and explicit classifications of inequivalent $k$-MUBs subsets for all prime $d\leq37$ (see Table~\ref{tab:MUB_Equiv.Classes} ).

 %Taking $d=5$ as an example, the finite unitary transformation set is $\mathcal{U}_{\text{finite}} = \{M_0, \dots, M_5\}$. By sequentially applying each unitary transformation $U \in \mathcal{U}_{\mathrm{finite}}$ and its complex conjugate to the full MUBs set $\mathcal{M}$, the complete transformation table can be obtained (see Table~\ref{tab:mub_d5_transform}).

  %  In the $d=5$ dimensional space, when $k=2$, all 15 subsets formed by selecting any two bases from $\mathcal{M}$ can be mutually transformed according to the transformation table (Table~\ref{tab:mub_d5_transform}), confirming that all two-bases subsets belong to the same equivalence class. When $k=3$, two distinct equivalence classes can be precisely identified, corresponding to initial subsets such as $\{M_0, M_1, M_2\}$ and $\{M_0, M_1, M_3\}$ (see Appendix D for the detailed partition of subsets). This result is in complete agreement with previous studies \cite{}.

   % Using the above method, we further analyzed the upper bounds and specific classifications of inequivalent classes for subsets of $k$ MUBs in dimensions $d \leq 37$. Detailed results are presented in Table~\ref{tab:MUB_Equiv.Classes} and the Appendix.

    \textit{Lower bound of Inequivalent MUBs Classes Based on the Shannon Entropy Criterion.---} Previous studies on inequivalent MUBs have mainly provided lower bounds using criteria such as robustness of incompatible measurements~\cite{Designolle.PhysRevLett.122.050402} or the Shannon entropy~\cite{Laura.Phys.Rev.Res.7.033152}. Adopting the latter, we classified all subsets for $d \leq 17$ (Table~\ref{tab:MUB_Equiv.Classes}), which clearly distinguishes inequivalent classes in low dimensions (e.g., $d=5$, $k=3$, entropy sums $4.43$ vs.\ $4.64$). For higher $d$, this method  suffers from exponential complexity and diminishing distinguishability, which lead to intractability and numerical instability. So we applied a sampling strategy that confirms reliable classification, with inter-class gaps exceeding intra-class fluctuations by two orders of magnitude.

    \begin{table*}[htbp]
    	\centering
    	\captionsetup{
    		justification=justified,
    		singlelinecheck=false
    	}
    	\caption{\label{tab:MUB_Equiv.Classes}%
    	\justifying
      A comparison of the amount of inequivalence classes generated by various MUBs subsets classification methods.The $\mathbf{blue}$ values represent the amount  of classes identified using the proposed finite equivalence operational method.The $\mathbf{purple}$ values show the amount of classes derived from the Shannon entropy criterion.The $\mathbf{cyan}$ values indicate the amount of classes obtained using the robustness of incompatible measurements. Crucially, for dimensions up to $d \le 17$, the classification results from the finite operation method and the entropy criterion exactly coincide, thereby establishing the precise equivalence classification of MUBs subsets in this range.}
        \begin{ruledtabular}
        \begin{tabular}{l||cccccccccccc}
            k\textbackslash d & 5 & 7 & 8 & 9 & 11 & 13 & 17 & 19 & 23 & 29 & 31 & 37 \\
			\midrule\midrule
			3 & \textcolor{blue}{2}\textbar\textcolor{purple}{2}\textbar\textcolor{cyan}{2} & \textcolor{blue}{1}\textbar\textcolor{purple}{1}\textbar\textcolor{cyan}{1} & \textcolor{blue}{1}\textbar\textcolor{purple}{1}\textbar\textcolor{cyan}{1} & \textcolor{blue}{2}\textbar\textcolor{purple}{2}\textbar\textcolor{cyan}{2} & \textcolor{blue}{1}\textbar\textcolor{purple}{1}\textbar\textcolor{cyan}{1} & \textcolor{blue}{2}\textbar\textcolor{purple}{2}\textbar\textcolor{cyan}{2} & \textcolor{blue}{2}\textbar\textcolor{purple}{2}\textbar\textcolor{cyan}{2} & \textcolor{blue}{1}\textbar\textcolor{purple}{1}\textbar\textcolor{cyan}{1} & \textcolor{blue}{1}\textbar\textcolor{cyan}{1} & \textcolor{blue}{2}\textbar\textcolor{cyan}{2} & \textcolor{blue}{1}\textbar\textcolor{cyan}{1} & \textcolor{blue}{2} \\
			4 & \textcolor{blue}{1}\textbar\textcolor{purple}{1}\textbar\textcolor{cyan}{1} & \textcolor{blue}{2}\textbar\textcolor{purple}{2}\textbar\textcolor{cyan}{2} & \textcolor{blue}{1}\textbar\textcolor{purple}{1}\textbar\textcolor{cyan}{1} & \textcolor{blue}{3}\textbar\textcolor{purple}{3}\textbar\textcolor{cyan}{3} & \textcolor{blue}{2}\textbar\textcolor{purple}{2}\textbar\textcolor{cyan}{2} & \textcolor{blue}{4}\textbar\textcolor{purple}{4}\textbar\textcolor{cyan}{4} & \textcolor{blue}{4}\textbar\textcolor{purple}{4}\textbar\textcolor{cyan}{4} & \textcolor{blue}{4}\textbar\textcolor{purple}{4}\textbar\textcolor{cyan}{4} & \textcolor{blue}{4}\textbar\textcolor{cyan}{4} & \textcolor{blue}{6}\textbar\textcolor{cyan}{6} & \textcolor{blue}{6}\textbar\textcolor{cyan}{6} & \textcolor{blue}{9} \\
			5 & \textcolor{blue}{1}\textbar\textcolor{purple}{1}\textbar\textcolor{cyan}{1} & \textcolor{blue}{1}\textbar\textcolor{purple}{1}\textbar\textcolor{cyan}{1} & \textcolor{blue}{1}\textbar\textcolor{purple}{1}\textbar\textcolor{cyan}{1} & \textcolor{blue}{3}\textbar\textcolor{purple}{3}\textbar\textcolor{cyan}{3} & \textcolor{blue}{2}\textbar\textcolor{purple}{2}\textbar\textcolor{cyan}{2} & \textcolor{blue}{5}\textbar\textcolor{purple}{5}\textbar\textcolor{cyan}{5} & \textcolor{blue}{8}\textbar\textcolor{purple}{8}\textbar\textcolor{cyan}{8} & \textcolor{blue}{5}\textbar\textcolor{purple}{5}\textbar\textcolor{cyan}{5} & \textcolor{blue}{6}\textbar\textcolor{cyan}{6} & \textcolor{blue}{19}\textbar\textcolor{cyan}{19} & \textcolor{blue}{11}\textbar\textcolor{cyan}{11} & \textcolor{blue}{29} \\
			6 & \textcolor{blue}{1}\textbar\textcolor{purple}{1}\textbar\textcolor{cyan}{1} & \textcolor{blue}{1}\textbar\textcolor{purple}{1}\textbar\textcolor{cyan}{1} & \textcolor{blue}{1}\textbar\textcolor{purple}{1}\textbar\textcolor{cyan}{1} & \textcolor{blue}{3}\textbar\textcolor{purple}{3}\textbar\textcolor{cyan}{3} & \textcolor{blue}{4}\textbar\textcolor{purple}{4}\textbar\textcolor{cyan}{4} & \textcolor{blue}{7}\textbar\textcolor{purple}{7}\textbar\textcolor{cyan}{7} & \textcolor{blue}{15}\textbar\textcolor{purple}{15}\textbar\textcolor{cyan}{15} & \textcolor{blue}{13}\textbar\textcolor{purple}{13}\textbar\textcolor{cyan}{13} & \textcolor{blue}{22}\textbar\textcolor{cyan}{22} & \textcolor{blue}{68}\textbar\textcolor{cyan}{67} & \textcolor{blue}{51}\textbar\textcolor{cyan}{50} & \textcolor{blue}{140} \\
			7 & & \textcolor{blue}{1}\textbar\textcolor{purple}{1}\textbar\textcolor{cyan}{1} & \textcolor{blue}{1}\textbar\textcolor{purple}{1}\textbar\textcolor{cyan}{1} & \textcolor{blue}{2}\textbar\textcolor{purple}{2}\textbar\textcolor{cyan}{2} & \textcolor{blue}{2}\textbar\textcolor{purple}{2}\textbar\textcolor{cyan}{2} & \textcolor{blue}{10}\textbar\textcolor{purple}{10}\textbar\textcolor{cyan}{10} & \textcolor{blue}{20}\textbar\textcolor{purple}{20}\textbar\textcolor{cyan}{20} & \textcolor{blue}{18}\textbar\textcolor{purple}{18}\textbar\textcolor{cyan}{18} & \textcolor{blue}{36}\textbar\textcolor{cyan}{32} & \textcolor{blue}{194}\textbar\textcolor{cyan}{145} & \textcolor{blue}{132}\textbar\textcolor{cyan}{92} & \textcolor{blue}{552} \\
			8 & & \textcolor{blue}{1}\textbar\textcolor{purple}{1}\textbar\textcolor{cyan}{1} & \textcolor{blue}{1}\textbar\textcolor{purple}{1}\textbar\textcolor{cyan}{1} & \textcolor{blue}{1}\textbar\textcolor{purple}{1}\textbar\textcolor{cyan}{1} & \textcolor{blue}{2}\textbar\textcolor{purple}{2}\textbar\textcolor{cyan}{2} & \textcolor{blue}{7}\textbar\textcolor{purple}{7}\textbar\textcolor{cyan}{7} & \textcolor{blue}{27}\textbar\textcolor{purple}{27}\textbar\textcolor{cyan}{23} & \textcolor{blue}{31}\textbar\textcolor{cyan}{22} & \textcolor{blue}{83}\textbar\textcolor{cyan}{35} & \textcolor{blue}{531}\textbar\textcolor{cyan}{?} & \textcolor{blue}{415}\textbar\textcolor{cyan}{?} & \textcolor{blue}{2044} \\
			9 & & & \textcolor{blue}{1}\textbar\textcolor{purple}{1} & \textcolor{blue}{1}\textbar\textcolor{purple}{1} & \textcolor{blue}{1}\textbar\textcolor{purple}{1} & \textcolor{blue}{5}\textbar\textcolor{purple}{5} & \textcolor{blue}{34}\textbar\textcolor{purple}{34} & \textcolor{blue}{33} & \textcolor{blue}{125} & \textcolor{blue}{1255} & \textcolor{blue}{992} & \textcolor{blue}{6624} \\
			10 & & & & \textcolor{blue}{1}\textbar\textcolor{purple}{1} & \textcolor{blue}{1}\textbar\textcolor{purple}{1} & \textcolor{blue}{4}\textbar\textcolor{purple}{4} & \textcolor{blue}{27}\textbar\textcolor{purple}{27} & \textcolor{blue}{44} & \textcolor{blue}{196} & \textcolor{blue}{2576} & \textcolor{blue}{2318} & \\
			11 & & & & & \textcolor{blue}{1}\textbar\textcolor{purple}{1} & \textcolor{blue}{2}\textbar\textcolor{purple}{2} & \textcolor{blue}{20}\textbar\textcolor{purple}{20} & \textcolor{blue}{33} & \textcolor{blue}{227} & \textcolor{blue}{4628} & & \\
			12 & & & & & \textcolor{blue}{1}\textbar\textcolor{purple}{1} & \textcolor{blue}{1}\textbar\textcolor{purple}{1} & \textcolor{blue}{15}\textbar\textcolor{purple}{15} & \textcolor{blue}{31} & \textcolor{blue}{268} & & & \\
			13 & & & & & & \textcolor{blue}{1}\textbar\textcolor{purple}{1} & \textcolor{blue}{8}\textbar\textcolor{purple}{8} & \textcolor{blue}{18} & \textcolor{blue}{227} & & & \\
			14 & & & & & & \textcolor{blue}{1}\textbar\textcolor{purple}{1} & \textcolor{blue}{4}\textbar\textcolor{purple}{4} & \textcolor{blue}{13} & \textcolor{blue}{196} & & & \\
			15 & & & & & & & \textcolor{blue}{2}\textbar\textcolor{purple}{2} & \textcolor{blue}{5} & \textcolor{blue}{125} & & & \\
			16 & & & & & & & \textcolor{blue}{1}\textbar\textcolor{purple}{1} & \textcolor{blue}{4} & \textcolor{blue}{83} & & & \\
			17 & & & & & & & \textcolor{blue}{1}\textbar\textcolor{purple}{1} & \textcolor{blue}{1} & \textcolor{blue}{36} & & & \\
			18 & & & & & & & \textcolor{blue}{1}\textbar\textcolor{purple}{1} & \textcolor{blue}{1} & \textcolor{blue}{22} & & & \\
			19 & & & & & & & & \textcolor{blue}{1} & \textcolor{blue}{6} & & & \\
			20 & & & & & & & & \textcolor{blue}{1} & \textcolor{blue}{5} & & & \\
            \end{tabular}
        \end{ruledtabular}
    \end{table*}

    \textit{Exact Classification of MUBs for Prime Dimensions $d\leq 17$.---} As shown in Table~\ref{tab:MUB_Equiv.Classes}, we compared the upper bounds on the amount of inequivalent MUBs classes obtained via the finite-operation transformation method with the lower bounds from the Shannon entropy criterion. Remarkably, for all $d \leq 17$, the two bounds coincide exactly, yielding the first complete classification of inequivalent MUBs subsets in this regime. Moreover, our upper bounds results exhibit strong consistency with the lower bounds obtained from the robustness of incompatible measurements criterion (covering partial cases with $d \leq 31$ and $3\leq k\leq 8$)~\cite{Designolle.PhysRevLett.122.050402}, with only one exception in all $d \leq 17$  : in the case $d=17, k=8$, our method gives 27 classes (matching the Shannon entropy entropy criterion) while robustness yields 23. This discrepancy highlights a key advantage of our approach: entropy- and robustness-based criteria are sensitive to numerical precision and may merge nearly indistinguishable classes (with entropy differences $<0.01$), whereas our transformation-based method remains virtually unaffected by such errors. Detailed classifications are provided in Sec.VII of SM. These features make our approach particularly powerful for reliably distinguishing inequivalent MUBs subsets, especially in higher dimensions.

   In higher dimensions ($d \ge 19$), we provided  upper bounds on the amount of inequivalent MUBs subsets up to $d=37$ (Table~\ref{tab:MUB_Equiv.Classes}). When the subset size $k$ is relatively small (e.g., $k \le 5$), these upper bounds exactly coincide with the lower bounds derived from the robustness of incompatible measurements criterion. This congruence signifies that the exact amount of inequivalent MUBs subsets is precisely determined for these low-order subsets. However, for larger values of $k$, some discrepancies emerge between our derived upper bounds and the known lower bounds. In these instances, the exact amount of inequivalent classes is strictly constrained between our calculated upper bounds and the previously established lower bounds.

    \textit{Distinguishing Inequivalent MUBs Subsets in Prime-Power Dimensions.---} This finite-operation MUBs classification method can be extended to Hilbert spaces of prime-power dimension $d = p^n$~\cite{William.Ann.Phys.191.363,Chaturvedi.PhysRevA.65.044301} to explore their MUBs subset inequivalence classes. In prime-power dimensions, constructing the candidate set of unitary transformations requires the inclusion of equivalent $\text{MUBs}$ generated by column-permutation operations $P$ ($P P^\mathsf{T} = I$), specifically $M_x P$, which reorder the column vectors within each basis. We rigorously demonstrate that it suffices to analyze the transformation paths of $M_d P$ (i.e., the action of the permutation matrix $P$ itself) to uniquely determine all equivalent transformations $M_x P$ (Sec.VI of SM). Therefore, the extended set of unitary transformations is defined as:
\begin{equation}
    \mathcal{U}_{\mathrm{Efinite}} = \left\{ U \in \mathcal{U}_{\mathrm{finite}}\right\} \cup \left\{ P \in \mathcal{P}_d \right\},
\end{equation}
where $\mathcal{P}_d$ is the set of column permutations that preserve the complete transformation property. While $d!$ permutations exist in total, only those maintaining MUBs completeness are included (constructive method in Sec.V of SM).
Notably, in prime dimensions ($d=p$), the inclusion of column permutations does not fundamentally alter the classification, simplifying $\mathcal{U}_{\mathrm{Efinite}}$ back to $\mathcal{U}_{\mathrm{finite}}$. Finally, the full finite-operation set, composed of the extended unitary transformations $\mathcal{U}_{\mathrm{Efinite}}$ and complex conjugation, allows us to apply the complete classification method to study inequivalent MUBs subsets in all prime-power dimensions.

   We applied the extended finite-operation classification method to the prime-power dimensions ($d=p^n$), based on the maximal complete $\text{MUBs}$ constructions of Ioannou \textit{et al.}~\cite{Ioannou.PhysRevLett.129.190401}. For $d=8$ ($p=2, n=3$) with 9 maximal $\text{MUBs}$, $168 = 3 \times 8 \times 7$ column permutations preserve completeness, and all $k$-base subsets form a single equivalence class (Table~\ref{tab:MUB_Equiv.Classes}), consistent with lower bounds from Shannon entropy and robustness of incompatible measurements. In $d=9$ ($p=3, n=2$), our upper bounds exactly match these lower bounds, e.g., $k=4$ yields three inequivalent classes. We further extended the study to $d = 16$ (Sec.VII of SM), demonstrating the broad applicability and reliability of our method in prime-power hilbert spaces.

   \textit{ Resource Consumption Analysis.---} Evaluating the effectiveness of MUBs subset classification naturally requires considering its computational cost. Here, we compare the three methods by analyzing the resources needed to distingwish inequivalence classes of $k$-MUBs subsets. A detailed discussion is provided in Sec.~VIII of the SM.
Firstly, the finite-operation method reduces the classification of MUBs subsets to a discrete transformation-matching problem. A $[(d+2)\times(d+1)]$ transformation table, built from the actions of $\mathcal{U}_{\mathrm{finite}}$ and complex conjugation, can be generated efficiently and quickly. For each $k$-subset check, extracting and normalizing the corresponding columns requires a complexity of $\mathcal{O}((d+2)\, k \log k)$. Since this procedure must be repeated $\binom{d+1}{k}$ times, the total complexity is
\begin{equation}
	T_U = \binom{d+1}{k} \cdot (d+2) \cdot k \log k \cdot \mathcal{O}(d,\text{other}).
\end{equation}
%The method avoids any high-dimensional optimization, providing a significant computational advantage.
where $\mathcal{O}(d,\text{other})$ denotes the fixed overhead associated with extracting and storing equivalent subsets. Secondly, the Shannon-entropy-based method, in principle, requires exploring the entire pure states in the Hilbert space to determine the minimal entropy sum.  For each $k$-subset check, a $d$-dimensional pure state $|\psi\rangle$ can be parameterized by $2d-1$ parameters ($d-1$ amplitudes $\alpha_m \in [-\pi,\pi]$ and $d$ phases $\phi_n \in [0,2\pi]$), and if each parameter is discretized into $s$ points, a total of $s^{2d-1}$ states must be evaluated. Consequently, the total complexity of this method can be easily characterized as \begin{equation}
	T_S = \binom{d+1}{k} \cdot s^{2d-1} \cdot \mathcal{O}(k,d,\text{other}).
\end{equation}
where $\mathcal{O}(k,d,\text{other})$ depend on the cost of single measurement and entropy calculation. Thirdly, the classification based on the robustness of measurement incompatibility requires solving a semidefinite program (SDP) for each of the $\binom{d+1}{k}$ subsets.  For a given subset, the SDP involves $d^{k}$ classical strategies, $(d^{2}-1)$ variables per operator $G_\lambda$, and on the order of $dk$ constraints, with parameter scans contributing only subdominant overhead. The resulting total complexity scales as
\begin{equation}
	T_R = \binom{d+1}{k} \cdot d^k \cdot (d^2-1) \cdot d\cdot k \cdot \mathcal{O}(s,\lambda,\text{other}).
\end{equation}
where $\mathcal{O}(s,\lambda, \text{other})$ accounts for the SDP-solving overhead. 

\begin{figure}[htbp]
    	\centering
        \includegraphics[width=0.9\linewidth]{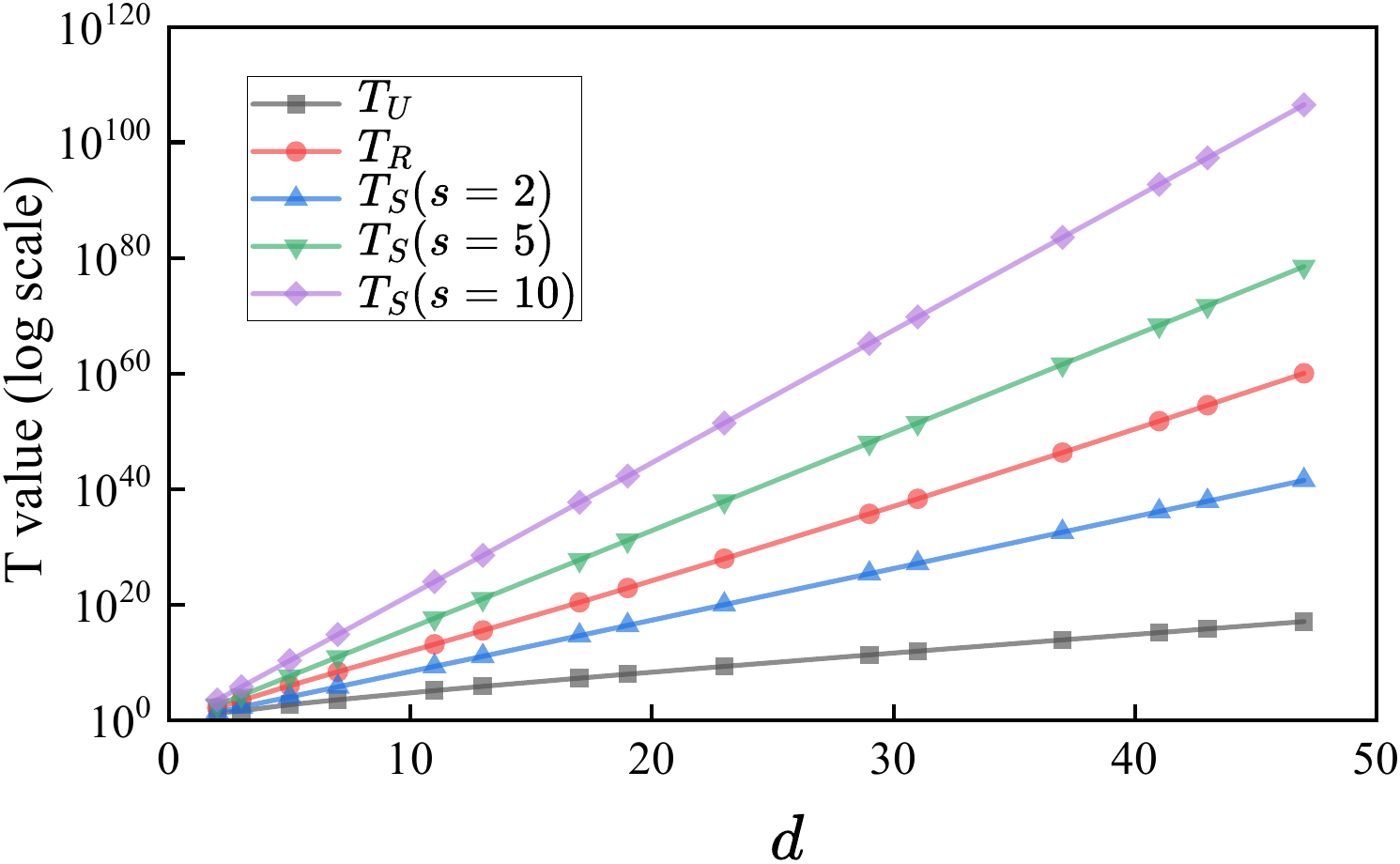}
    	\caption{\justifying
       Comparison of computational complexity rates as a function of the dimension $d$ for the classification of MUB subsets with subset size $k=(d+1)/2$, based on Shannon entropy, robustness of measurement incompatibility and finite equivalence operations, under different sampling densities $s=2,5,10$.}
    	\label{fig:comput_complexity}
    \end{figure}

As shown in Fig.~\ref{fig:comput_complexity}, the computational complexities of the three classification methods can be compared intuitively. For a fixed subset size $k$, the figure illustrates how $T_S$, $T_R$, and $T_U$ scale with the dimension $d$. The Shannon-entropy-based method requires exponential sampling ($T_S \propto s^{2d-1}$), leading to an exponential increase in complexity with $d$. The classification based on the robustness of measurement incompatibility similarly exhibits exponential growth with $d$ and $k$, due to the need to solve a large semidefinite program (SDP) for each subset. In contrast, the finite-operation method scales only linearly with $d$, avoiding high-dimensional numerical optimization and significantly improving computational efficiency.

%Fig.~\ref{fig:comput_complexity} compares the computational complexity of the two methods across different sampling densities. For a fixed subset size $k$, it shows how $T_S$ and $T_U$ scale with the dimension $d$. Shannon entropy criterion requires exponential sampling ($T_S \propto s^{2d-1}$), causing its complexity to grow exponentially with $d$. In contrast, our finite-operation transformation method reduces classification to a discrete search: the main bottleneck arises from the combinatorial factor $\binom{d+1}{k}$, while the core transformation operations scale only linearly with $d$, fully avoiding the exponential overhead of high-dimensional numerical optimization.

In Sec.IX of SM, we further evaluate computational time of the other MUBs subsets classification approaches—including numerical optimization based on robustness of incompatible measurements, steering robustness, QRAC accuracy, and diamond distance. It is shown that the finite-operation method drastically reduces both computation time and cost while preserving accuracy. For instance, for $d=31, k=8$, competing methods approach their processing limits, whereas our method completes full classification in under three minutes.

    %Figure~\ref{fig:comput_complexity} illustrates the comparison of computational complexity between the two methods under different sampling densities. For a fixed subset size $k$, the trends of $T_S$ and $T_U$ as the dimension $d$ increases are shown. The sum of Shannon entropies method requires exponential sampling ($T_S \propto s^{2d-1}$), causing its total computational complexity to grow exponentially with $d$. In contrast, the finite operation transformation method proposed in this study transforms the classification problem into a discrete search, with the main complexity bottleneck determined by the combinatorial number $\binom{d+1}{k}$, while the complexity of the core transformation operations grows only linearly with $d$, completely avoiding the exponential bottleneck of high-dimensional numerical optimization.

   % In the Appendix H, we further analyze and evaluate other MUBs subset equivalence classification methods, including numerical optimization strategies based on joint-measurement robustness, Steering robustness, the QRAC criterion, and the Diamond distance criterion, in terms of computational time. The results indicate that the finite equivalence operation classification method significantly reduces computation time and cost while maintaining accuracy. For example, in the task of distinguishing MUBs inequivalent classes for $d=31, k=8$, other methods are nearly at their processing limit, whereas this method completes the full classification in less than 3 minutes.

    \textit{Conclusions and Discussions.---} %This work establishes a systematic framework for the inequivalent classification of subsets of maximal complete mutually unbiased bases (MUBs) in Hilbert space. First, we presented the mapping solutions of MUBs in prime dimensions and provided an analytical upper bound $\binom{d+1}{k}/2d$ on the number of inequivalent subsets. By further incorporating complex conjugation as an equivalence operation, tighter upper bounds were obtained for $d \le 37$. Using the sum-of-Shannon-entropies and minimum criterion to compute lower bounds, we determined for the first time the exact classification of inequivalent MUB subsets for $d \le 17$. To extend the applicability of the method, we generalized the domain of unitary transformations to include equivalence MUBs obtained via column permutations of the bases, enabling subset classification in prime-power-dimensional Hilbert spaces. Furthermore, a comparison of computational resource consumption between the finite equivalence operations method and the sum-of-Shannon-entropies approach demonstrates that our method significantly reduces computational cost. Future research will focus on the classification of incomplete MUB sets, exploring their equivalence relations and classification strategies to broaden the applicability of mutually unbiased basis theory. Meanwhile, this classification framework has potential applications in quantum tomography, quantum cryptography, measurement design, and quantum error correction, providing theoretical and algorithmic support for the analysis of high-dimensional quantum systems and quantum resource certification.
   In conclusion, we have introduced a robust and resource-efficient finite-operation framework for the structural classification of MUBs subsets. For arbitry sets of $k$ MUBs in any prime dimension $d$, this method yields a universal analytical upper bound for the amount of MUBs equivalence classes. By applying this method through simple iterations, we further obtain tighter classification upper bounds for prime dimensions $d\leq 37$. Crucially,  comparing these upper bounds with existing lower bounds, we successfully determine the exact inequivalence classes for all MUBs subsets in dimensions $d \leq 17$. Furthermore, the framework's extension to prime-power dimensions and its proven computational efficiency mark a significant step forward. This systematic classification provides robust theoretical and algorithmic support essential for high-dimensional quantum resource certification and measurement design. In addition, the framework can be further tightened in prime-power dimensions by incorporating an overall global phase into the unitary operation~$\mathcal{U}$. More broadly, the classification method developed here is directly useful for a variety of high-dimensional quantum information tasks, such as quantum randomness generation, quantum key distribution, multi-party cryptography, entanglement witnessing, quantum metrology, etc.

	\textit{Acknowledgment.---}C.R. was supported by the National Natural Science Foundation of China (Grant No. 12575016, 12421005,12247105), Hunan provincial major sci-tech program (No. 2023ZJ1010),  the Foundation Xiangjiang Laboratory (XJ2302001) and Xiaoxiang Scholars Program of Hunan Normal University. Z.P.X. acknowledges the support from {National Natural Science Foundation of China} (Grant No.~12305007), 
Anhui Provincial Natural Science Foundation (Grant No.~2308085QA29, No.~2508085Y003), Anhui Province
Science and Technology Innovation Project (No.~202423r06050004). 
	
	%==============================================================================================
	\bibliographystyle{apsrev4-2}
	\bibliography{references}

@article{Schwinger.pnas.46.4.570,
	author = {Julian Schwinger },
	title = {UNITARY OPERATOR BASES<sup>*</sup>},
	journal = {Proc. Natl. Acad. Sci.},
	volume = {46},
	number = {4},
	pages = {570-579},
	year = {1960},
	doi = {10.1073/pnas.46.4.570}
}

@article{NicolasJ.PhysRevLett.88.127902,
	title = {Security of Quantum Key Distribution Using $\mathit{d}$-Level Systems},
	author = {Cerf, Nicolas J. and Bourennane, Mohamed and Karlsson, Anders and Gisin, Nicolas},
	journal = {Phys. Rev. Lett.},
	volume = {88},
	issue = {12},
	pages = {127902},
	numpages = {4},
	year = {2002},
	month = {Mar},
	publisher = {American Physical Society},
	doi = {10.1103/PhysRevLett.88.127902}
}

@article{Uola.RevModPhys.95.011003,
	title = {Colloquium: Incompatible measurements in quantum information science},
	author = {G\"uhne, Otfried and Haapasalo, Erkka and Kraft, Tristan and Pellonp\"a\"a, Juha-Pekka and Uola, Roope},
	journal = {Rev. Mod. Phys.},
	volume = {95},
	issue = {1},
	pages = {011003},
	numpages = {25},
	year = {2023},
	month = {Feb},
	publisher = {American Physical Society},
	doi = {10.1103/RevModPhys.95.011003}
}

@article{Bendersky.PhysRevLett.100.190403,
	title = {Selective and Efficient Estimation of Parameters for Quantum Process Tomography},
	author = {Bendersky, Ariel and Pastawski, Fernando and Paz, Juan Pablo},
	journal = {Phys. Rev. Lett.},
	volume = {100},
	issue = {19},
	pages = {190403},
	numpages = {4},
	year = {2008},
	month = {May},
	publisher = {American Physical Society},
	doi = {10.1103/PhysRevLett.100.190403}
}

@article{Spengler.PhysRevA.86.022311,
	title = {Entanglement detection via mutually unbiased bases},
	author = {Spengler, Christoph and Huber, Marcus and Brierley, Stephen and Adaktylos, Theodor and Hiesmayr, Beatrix C.},
	journal = {Phys. Rev. A},
	volume = {86},
	issue = {2},
	pages = {022311},
	numpages = {8},
	year = {2012},
	month = {Aug},
	publisher = {American Physical Society},
	doi = {10.1103/PhysRevA.86.022311}
}

@article{Aharonov.PhysRevLett.47.1029,
	title = {New Interpretation of the Scalar Product in Hilbert Space},
	author = {Aharonov, Y. and Albert, David Z. and Au, C. K.},
	journal = {Phys. Rev. Lett.},
	volume = {47},
	issue = {15},
	pages = {1029--1031},
	numpages = {0},
	year = {1981},
	month = {Oct},
	publisher = {American Physical Society},
	doi = {10.1103/PhysRevLett.47.1029}
}

@article{Fano.Rev.Mod.Phys.29.74,
	title = {Description of States in Quantum Mechanics by Density Matrix and Operator Techniques},
	author = {Fano, U.},
	journal = {Rev. Mod. Phys.},
	volume = {29},
	issue = {1},
	pages = {74--93},
	numpages = {0},
	year = {1957},
	month = {Jan},
	publisher = {American Physical Society},
	doi = {10.1103/RevModPhys.29.74}
}

@article{Siudzińska.Sci.Rep.11.22988,
	title = {Entanglement witnesses from mutually unbiased measurements},
	author = {Siudzińska, Katarzyna and Chruściński, Dariusz},
	journal = {Sci. Rep.},
	volume = {11},
	pages = {22988},
	year = {2021},
	doi = {10.1038/s41598-021-02356-2}
}

@article{Ivonovic.J.Phys.A.14.3241,
	title = {Geometrical description of quantal state determination},
	author = {I D Ivonovic},
	journal = {J. Phys. A},
	volume = {14},
	pages = {3241},
	year = {1981},
	doi = {10.1088/0305-4470/14/12/019}
}

@article{William.Ann.Phys.191.363,
	title = {Optimal state-determination by mutually unbiased measurements},
	journal = {Ann. Phys. (N.Y.)},
	volume = {191},
	number = {2},
	pages = {363},
	year = {1989},
	issn = {0003-4916},
	doi = {https://doi.org/10.1016/0003-4916(89)90322-9},
	author = {William K Wootters and Brian D Fields}
}

@article{Parthasarathy.QPRT.7,
	author = {Parthasarathy, K.},
	year = {2004},
	month = {09},
	pages = {},
	title = {On Estimating the State of a Finite Level Quantum System},
	volume = {7},
	journal = {Infin. Dimens. Ana.,Quantum Probab. Relat. Top.},
	doi = {10.1142/S0219025704001797}
}

@article{Adamson.PhysRevLett.105.030406,
	title = {Improving Quantum State Estimation with Mutually Unbiased Bases},
	author = {Adamson, R. B. A. and Steinberg, A. M.},
	journal = {Phys. Rev. Lett.},
	volume = {105},
	issue = {3},
	pages = {030406},
	numpages = {4},
	year = {2010},
	month = {Jul},
	publisher = {American Physical Society},
	doi = {10.1103/PhysRevLett.105.030406}
}

@article{Jaroslav.PhysRevA.92.052303,
	title = {Least-bias state estimation with incomplete unbiased measurements},
	author = {\ifmmode \check{R}\else \v{R}\fi{}eh\'a\ifmmode \check{c}\else \v{c}\fi{}ek, Jaroslav and Hradil, Zden\ifmmode \check{e}\else \v{e}\fi{}k and Teo, Yong Siah and S\'anchez-Soto, Luis L. and Ng, Hui Khoon and Chai, Jing Hao and Englert, Berthold-Georg},
	journal = {Phys. Rev. A},
	volume = {92},
	issue = {5},
	pages = {052303},
	numpages = {13},
	year = {2015},
	month = {Nov},
	publisher = {American Physical Society},
	doi = {10.1103/PhysRevA.92.052303}
}

@article{Liu.Sci.Rep.5.13138,
	title = {Separability criteria via sets of mutually unbiased measurements},
	author = {Liu, Lu and Gao, Ting and Yan, Fengli},
	journal = {Sci. Rep.},
	volume = {5},
	pages = {13138},
	year = {2015},
	doi = {10.1038/srep13138}
}

@article{Giovannini.PhysRevLett.110.143601,
	title = {Characterization of High-Dimensional Entangled Systems via Mutually Unbiased Measurements},
	author = {Giovannini, D. and Romero, J. and Leach, J. and Dudley, A. and Forbes, A. and Padgett, M. J.},
	journal = {Phys. Rev. Lett.},
	volume = {110},
	issue = {14},
	pages = {143601},
	numpages = {5},
	year = {2013},
	month = {Apr},
	publisher = {American Physical Society},
	doi = {10.1103/PhysRevLett.110.143601}
}

@article{Sosa.PhysRevLett.119.150401,
	title = {Experimental Study of Optimal Measurements for Quantum State Tomography},
	author = {Sosa-Martinez, H. and Lysne, N. K. and Baldwin, C. H. and Kalev, A. and Deutsch, I. H. and Jessen, P. S.},
	journal = {Phys. Rev. Lett.},
	volume = {119},
	issue = {15},
	pages = {150401},
	numpages = {6},
	year = {2017},
	month = {Oct},
	publisher = {American Physical Society},
	doi = {10.1103/PhysRevLett.119.150401}
}

@article{Fern.PhysRevA.83.052332,
	title = {Quantum process reconstruction based on mutually unbiased basis},
	author = {Fern\'andez-P\'erez, A. and Klimov, A. B. and Saavedra, C.},
	journal = {Phys. Rev. A},
	volume = {83},
	issue = {5},
	pages = {052332},
	numpages = {6},
	year = {2011},
	month = {May},
	publisher = {American Physical Society},
	doi = {10.1103/PhysRevA.83.052332}
}

@article{Armin.Sci.Adv.7.eabc3847,
	author = {Armin Tavakoli  and Máté Farkas  and Denis Rosset  and Jean-Daniel Bancal  and Jedrzej Kaniewski },
	title = {Mutually unbiased bases and symmetric informationally complete measurements in Bell experiments},
	journal = {Sci. Adv.},
	volume = {7},
	number = {7},
	pages = {eabc3847},
	year = {2021},
	doi = {10.1126/sciadv.abc3847}
}

@article{Brierley.QuantumInf.Comput.10.0803,
	title = {All Mutually Unbiased Bases in Dimensions Two to Five},
	author = {Brierley, Stephen and Weigert, Stefan and Bengtsson, Ingemar},
	journal = {Quantum Inf. Comput.},
	volume = {10},
	pages = {0803},
	year = {2010},
	month = {Jul},
	doi = {10.26421/QIC10.9-10-6}
}

@article{Kraus.PhysRevD.35.3070,
	title = {Complementary observables and uncertainty relations},
	author = {Kraus, K.},
	journal = {Phys. Rev. D},
	volume = {35},
	issue = {10},
	pages = {3070--3075},
	numpages = {0},
	year = {1987},
	month = {May},
	publisher = {American Physical Society},
	doi = {10.1103/PhysRevD.35.3070}
}

@article{Czartowski.PhysRevLett.124.090503,
	title = {Isoentangled Mutually Unbiased Bases, Symmetric Quantum Measurements, and Mixed-State Designs},
	author = {Czartowski, Jakub and Goyeneche, Dardo and Grassl, Markus and \ifmmode \dot{Z}\else \.{Z}\fi{}yczkowski, Karol},
	journal = {Phys. Rev. Lett.},
	volume = {124},
	issue = {9},
	pages = {090503},
	numpages = {6},
	year = {2020},
	month = {Mar},
	publisher = {American Physical Society},
	doi = {10.1103/PhysRevLett.124.090503}
}

@article{Yan.PhysRevLett.132.080202,
	title = {Experimental Demonstration of Inequivalent Mutually Unbiased Bases},
	author = {Yan, Wen-Zhe and Li, Yunting and Hou, Zhibo and Zhu, Huangjun and Xiang, Guo-Yong and Li, Chuan-Feng and Guo, Guang-Can},
	journal = {Phys. Rev. Lett.},
	volume = {132},
	issue = {8},
	pages = {080202},
	numpages = {7},
	year = {2024},
	month = {Feb},
	publisher = {American Physical Society},
	doi = {10.1103/PhysRevLett.132.080202}
}

@article{Deutsch.PhysRevLett.50.631,
	title = {Uncertainty in Quantum Measurements},
	author = {Deutsch, David},
	journal = {Phys. Rev. Lett.},
	volume = {50},
	issue = {9},
	pages = {631--633},
	numpages = {0},
	year = {1983},
	month = {Feb},
	publisher = {American Physical Society},
	doi = {10.1103/PhysRevLett.50.631}
}

@article{Maassen.PhysRevLett.60.1103,
	title = {Generalized entropic uncertainty relations},
	author = {Maassen, Hans and Uffink, J. B. M.},
	journal = {Phys. Rev. Lett.},
	volume = {60},
	issue = {12},
	pages = {1103--1106},
	numpages = {0},
	year = {1988},
	month = {Mar},
	publisher = {American Physical Society},
	doi = {10.1103/PhysRevLett.60.1103}
}

@article{Riccardi.PhysRevA.95.032109,
	title = {Tight entropic uncertainty relations for systems with dimension three to five},
	author = {Riccardi, Alberto and Macchiavello, Chiara and Maccone, Lorenzo},
	journal = {Phys. Rev. A},
	volume = {95},
	issue = {3},
	pages = {032109},
	numpages = {7},
	year = {2017},
	month = {Mar},
	publisher = {American Physical Society},
	doi = {10.1103/PhysRevA.95.032109}
}

@article{Trifonov.J.Phys.A.31.8041,
	doi = {10.1088/0305-4470/31/39/016},
	year = {1998},
	month = {oct},
	publisher = {},
	volume = {31},
	number = {39},
	pages = {8041},
	author = {D A Trifonov and S G Donev},
	title = {Characteristic uncertainty relations},
	journal = {J. Phys. A: Math. Gen.}
}

@article{Laura.Phys.Rev.Res.7.033152,
  title = {Complementarity-based complementarity: The choice of mutually unbiased observables shapes quantum uncertainty relations},
  author = {Serino, Laura and Chesi, Giovanni and Brecht, Benjamin and Maccone, Lorenzo and Macchiavello, Chiara and Silberhorn, Christine},
  journal = {Phys. Rev. Res.},
  volume = {7},
  issue = {3},
  pages = {033152},
  numpages = {9},
  year = {2025},
  month = {Aug},
  publisher = {American Physical Society},
  doi = {10.1103/v24q-sl6n}
}

@article{Wehner.New.J.Phys.12.025009,
	doi = {10.1088/1367-2630/12/2/025009},
	year = {2010},
	month = {feb},
	volume = {12},
	number = {2},
	pages = {025009},
	author = {Wehner, Stephanie and Winter, Andreas},
	title = {Entropic uncertainty relations—a survey},
	journal = {New J. Phys.}
}

@article{Coles.RevModPhys.89.015002,
	title = {Entropic uncertainty relations and their applications},
	author = {Coles, Patrick J. and Berta, Mario and Tomamichel, Marco and Wehner, Stephanie},
	journal = {Rev. Mod. Phys.},
	volume = {89},
	issue = {1},
	pages = {015002},
	numpages = {58},
	year = {2017},
	month = {Feb},
	publisher = {American Physical Society},
	doi = {10.1103/RevModPhys.89.015002}
}

@article{Designolle.PhysRevLett.122.050402,
	title = {Quantifying Measurement Incompatibility of Mutually Unbiased Bases},
	author = {Designolle, S\'ebastien and Skrzypczyk, Paul and Fr\"owis, Florian and Brunner, Nicolas},
	journal = {Phys. Rev. Lett.},
	volume = {122},
	issue = {5},
	pages = {050402},
	numpages = {6},
	year = {2019},
	month = {Feb},
	publisher = {American Physical Society},
	doi = {10.1103/PhysRevLett.122.050402}
}

@article{Designolle.PhysRevA.105.032430,
	title = {Robust genuine high-dimensional steering with many measurements},
	author = {Designolle, S\'ebastien},
	journal = {Phys. Rev. A},
	volume = {105},
	issue = {3},
	pages = {032430},
	numpages = {8},
	year = {2022},
	month = {Mar},
	publisher = {American Physical Society},
	doi = {10.1103/PhysRevA.105.032430}
}

@article{Aguilar.PhysRevLett.121.050501,
	title = {Connections between Mutually Unbiased Bases and Quantum Random Access Codes},
	author = {Aguilar, Edgar A. and Borka\l{}a, Jakub J. and Mironowicz, Piotr and Paw\l{}owski, Marcin},
	journal = {Phys. Rev. Lett.},
	volume = {121},
	issue = {5},
	pages = {050501},
	numpages = {6},
	year = {2018},
	month = {Jul},
	publisher = {American Physical Society},
	doi = {10.1103/PhysRevLett.121.050501}
}

@article{Hiesmayr.New.J.Phys.23.093018,
	title = {Detecting entanglement can be more effective with inequivalent mutually unbiased bases},
	volume = {23},
	issn = {1367-2630},
	doi = {10.1088/1367-2630/ac20ea},
	language = {en},
	number = {9},
	urldate = {2025-02-18},
	journal = {New J. Phys.},
	author = {Hiesmayr, B C and McNulty, D and Baek, S and Singha Roy, S and Bae, J and Chruściński, D},
	month = sep,
	year = {2021},
	pages = {093018}
}

@article{Tendick.PhysRevLett.131.120202,
	title = {Distributed Quantum Incompatibility},
	author = {Tendick, Lucas and Kampermann, Hermann and Bru\ss{}, Dagmar},
	journal = {Phys. Rev. Lett.},
	volume = {131},
	issue = {12},
	pages = {120202},
	numpages = {6},
	year = {2023},
	month = {Sep},
	publisher = {American Physical Society},
	doi = {10.1103/PhysRevLett.131.120202}
}

@misc{Thomas.arxiv.2207.05722,
      title={Quantifying the high-dimensionality of quantum devices}, 
      author={Thomas Cope and Roope Uola},
      year={2023},
      eprint={2207.05722},
      archivePrefix={arXiv},
      primaryClass={quant-ph},
     url={https://arxiv.org/abs/2207.05722}, 
}

@article{Tendick.Quantum.7.1003,
	title={Distance-based resource quantification for sets of quantum measurements},
	volume={7},
	ISSN={2521-327X},
	DOI={10.22331/q-2023-05-15-1003},
	journal={Quantum},
	publisher={Verein zur Forderung des Open Access Publizierens in den Quantenwissenschaften},
	author={Tendick, Lucas and Kliesch, Martin and Kampermann, Hermann and Bruß, Dagmar},
	year={2023},
	month=may, pages={1003} 
}

@article{Ioannou.PhysRevLett.129.190401,
	title = {Simulability of High-Dimensional Quantum Measurements},
	author = {Ioannou, Marie and Sekatski, Pavel and Designolle, S\'ebastien and Jones, Benjamin D. M. and Uola, Roope and Brunner, Nicolas},
	journal = {Phys. Rev. Lett.},
	volume = {129},
	issue = {19},
	pages = {190401},
	numpages = {7},
	year = {2022},
	month = {Nov},
	publisher = {American Physical Society},
	doi = {10.1103/PhysRevLett.129.190401}
}

@article{Cerf.PhysRevLett.88.127902,
	title = {Security of Quantum Key Distribution Using $\mathit{d}$-Level Systems},
	author = {Cerf, Nicolas J. and Bourennane, Mohamed and Karlsson, Anders and Gisin, Nicolas},
	journal = {Phys. Rev. Lett.},
	volume = {88},
	issue = {12},
	pages = {127902},
	numpages = {4},
	year = {2002},
	month = {Mar},
	publisher = {American Physical Society},
	doi = {10.1103/PhysRevLett.88.127902}
}

@article{Bechmann.PhysRevLett.85.3313,
	title = {Quantum Cryptography with 3-State Systems},
	author = {Bechmann-Pasquinucci, Helle and Peres, Asher},
	journal = {Phys. Rev. Lett.},
	volume = {85},
	issue = {15},
	pages = {3313--3316},
	numpages = {0},
	year = {2000},
	month = {Oct},
	publisher = {American Physical Society},
	doi = {10.1103/PhysRevLett.85.3313}
}

@article{Bent.PhysRevX.5.041006,
	title = {Experimental Realization of Quantum Tomography of Photonic Qudits via Symmetric Informationally Complete Positive Operator-Valued Measures},
	author = {Bent, N. and Qassim, H. and Tahir, A. A. and Sych, D. and Leuchs, G. and S\'anchez-Soto, L. L. and Karimi, E. and Boyd, R. W.},
	journal = {Phys. Rev. X},
	volume = {5},
	issue = {4},
	pages = {041006},
	numpages = {12},
	year = {2015},
	month = {Oct},
	publisher = {American Physical Society},
	doi = {10.1103/PhysRevX.5.041006}
}

@misc{Daniel.arxiv.2410.23997,
	title={Mutually Unbiased Bases in Composite Dimensions -- A Review}, 
	author={Daniel McNulty and Stefan Weigert},
	year={2024},
	eprint={2410.23997},
	archivePrefix={arXiv},
	primaryClass={quant-ph},
	url={https://arxiv.org/abs/2410.23997}
}

@article{Cao.PhysRevX.6.011020,
	title = {Source-Independent Quantum Random Number Generation},
	author = {Cao, Zhu and Zhou, Hongyi and Yuan, Xiao and Ma, Xiongfeng},
	journal = {Phys. Rev. X},
	volume = {6},
	issue = {1},
	pages = {011020},
	numpages = {11},
	year = {2016},
	month = {Feb},
	publisher = {American Physical Society},
	doi = {10.1103/PhysRevX.6.011020}
}

@article{Ma.PhysRevA.99.022328,
	title = {Coherence as a resource for source-independent quantum random-number generation},
	author = {Ma, Jiajun and Hakande, Aishwarya and Yuan, Xiao and Ma, Xiongfeng},
	journal = {Phys. Rev. A},
	volume = {99},
	issue = {2},
	pages = {022328},
	numpages = {12},
	year = {2019},
	month = {Feb},
	publisher = {American Physical Society},
	doi = {10.1103/PhysRevA.99.022328}
}

@article{Chaturvedi.PhysRevA.65.044301,
	title = {Aspects of mutually unbiased bases in odd-prime-power dimensions},
	author = {Chaturvedi, S.},
	journal = {Phys. Rev. A},
	volume = {65},
	issue = {4},
	pages = {044301},
	numpages = {3},
	year = {2002},
	month = {Mar},
	publisher = {American Physical Society},
	doi = {10.1103/PhysRevA.65.044301}
}

@article{Ikuta.PhysRevResearch.4.L042007,
	title = {Scalable implementation of $(d+1)$ mutually unbiased bases for $d$-dimensional quantum key distribution},
	author = {Ikuta, Takuya and Akibue, Seiseki and Yonezu, Yuya and Honjo, Toshimori and Takesue, Hiroki and Inoue, Kyo},
	journal = {Phys. Rev. Res.},
	volume = {4},
	issue = {4},
	pages = {L042007},
	numpages = {6},
	year = {2022},
	month = {Oct},
	publisher = {American Physical Society},
	doi = {10.1103/PhysRevResearch.4.L042007}
}

@article{Mafu.PhysRevA.88.032305,
	title = {Higher-dimensional orbital-angular-momentum-based quantum key distribution with mutually unbiased bases},
	author = {Mafu, Mhlambululi and Dudley, Angela and Goyal, Sandeep and Giovannini, Daniel and McLaren, Melanie and Padgett, Miles J. and Konrad, Thomas and Petruccione, Francesco and L\"utkenhaus, Norbert and Forbes, Andrew},
	journal = {Phys. Rev. A},
	volume = {88},
	issue = {3},
	pages = {032305},
	numpages = {8},
	year = {2013},
	month = {Sep},
	publisher = {American Physical Society},
	doi = {10.1103/PhysRevA.88.032305}
}

@article{Sheridan.PhysRevA.82.030301,
	title = {Security proof for quantum key distribution using qudit systems},
	author = {Sheridan, Lana and Scarani, Valerio},
	journal = {Phys. Rev. A},
	volume = {82},
	issue = {3},
	pages = {030301},
	numpages = {4},
	year = {2010},
	month = {Sep},
	publisher = {American Physical Society},
	doi = {10.1103/PhysRevA.82.030301}
}

@article{Wu.PhysRevA.79.022104,
	title = {Entropic uncertainty relation for mutually unbiased bases},
	author = {Wu, Shengjun and Yu, Sixia and M\o{}lmer, Klaus},
	journal = {Phys. Rev. A},
	volume = {79},
	issue = {2},
	pages = {022104},
	numpages = {5},
	year = {2009},
	month = {Feb},
	publisher = {American Physical Society},
	doi = {10.1103/PhysRevA.79.022104}
}

@article{Ballester.PhysRevA.75.022319,
	title = {Entropic uncertainty relations and locking: Tight bounds for mutually unbiased bases},
	author = {Ballester, Manuel A. and Wehner, Stephanie},
	journal = {Phys. Rev. A},
	volume = {75},
	issue = {2},
	pages = {022319},
	numpages = {8},
	year = {2007},
	month = {Feb},
	publisher = {American Physical Society},
	doi = {10.1103/PhysRevA.75.022319}
}

@article{Bartkiewicz.PhysRevA.93.062345,
	title = {Temporal steering and security of quantum key distribution with mutually unbiased bases against individual attacks},
	author = {Bartkiewicz, Karol and \ifmmode \check{C}\else \v{C}\fi{}ernoch, Anton\'{\i}n and Lemr, Karel and Miranowicz, Adam and Nori, Franco},
	journal = {Phys. Rev. A},
	volume = {93},
	issue = {6},
	pages = {062345},
	numpages = {7},
	year = {2016},
	month = {Jun},
	publisher = {American Physical Society},
	doi = {10.1103/PhysRevA.93.062345}
}

@article{Tavakoli.RevModPhys.96.045006,
  title = {Semidefinite programming relaxations for quantum correlations},
  author = {Tavakoli, Armin and Pozas-Kerstjens, Alejandro and Brown, Peter and Ara\'ujo, Mateus},
  journal = {Rev. Mod. Phys.},
  volume = {96},
  issue = {4},
  pages = {045006},
  numpages = {68},
  year = {2024},
  month = {Dec},
  publisher = {American Physical Society},
  doi = {10.1103/RevModPhys.96.045006}
}

@article{Kraus.PhysRevLett.104.020504,
	title = {Local Unitary Equivalence of Multipartite Pure States},
	author = {Kraus, B.},
	journal = {Phys. Rev. Lett.},
	volume = {104},
	issue = {2},
	pages = {020504},
	numpages = {4},
	year = {2010},
	month = {Jan},
	publisher = {American Physical Society},
	doi = {10.1103/PhysRevLett.104.020504}
}

@article{Kraus.PhysRevA.82.032121,
	title = {Local unitary equivalence and entanglement of multipartite pure states},
	author = {Kraus, B.},
	journal = {Phys. Rev. A},
	volume = {82},
	issue = {3},
	pages = {032121},
	numpages = {14},
	year = {2010},
	month = {Sep},
	publisher = {American Physical Society},
	doi = {10.1103/PhysRevA.82.032121}
}

@article{Lawrence.PhysRevA.65.032320,
	title = {Mutually unbiased binary observable sets on N qubits},
	author = {Lawrence, Jay and Brukner, \ifmmode \check{C}\else \v{C}\fi{}aslav and Zeilinger, Anton},
	journal = {Phys. Rev. A},
	volume = {65},
	issue = {3},
	pages = {032320},
	numpages = {5},
	year = {2002},
	month = {Feb},
	publisher = {American Physical Society},
	doi = {10.1103/PhysRevA.65.032320}
}

@article{Ambrosio.Sci.Rep.3.2726,
	title = {Test of mutually unbiased bases for six-dimensional photonic quantum systems},
	author = {D'Ambrosio, Vincenzo and Cardano, Filippo and Karimi, Ebrahim and Nagali, Eleonora and Santamato, Enrico and Marrucci, Lorenzo and Sciarrino, Fabio},
	journal = {Sci. Rep.},
	volume = {3},
	pages = {2726},
	year = {2013},
	doi = {10.1038/srep02726}
}

@article{Philippe.Crypt.Comm.2.211,
	title = {The problem of mutually unbiased bases in dimension 6},
	author = {Jaming,Philippe and Matolcsi,Máté and Móra,Péter},
	journal = {Crypt. Comm.},
	volume = {2},
	pages = {211},
	year = {2010},
	doi = {https://doi.org/10.1007/s12095-010-0023-1}
}

@article{Durt.J.Phys.A.38.5267,
	title = {About mutually unbiased bases in even and odd prime power dimensions},
	author = {Durt, Thomas},
	journal = {J. Phys. A},
	volume = {38},
	pages = {5267},
	year = {2005},
	month = {May},
	doi = {10.1088/0305-4470/38/23/013}
}

@article{Brierley.PhysRevA.79.052316,
	title = {Constructing mutually unbiased bases in dimension six},
	author = {Brierley, Stephen and Weigert, Stefan},
	journal = {Phys. Rev. A},
	volume = {79},
	issue = {5},
	pages = {052316},
	numpages = {13},
	year = {2009},
	month = {May},
	publisher = {American Physical Society},
	doi = {10.1103/PhysRevA.79.052316}
}

\end{document}